\documentclass[sigconf,natbib=true]{acmart}
\AtBeginDocument{%
  }

\setcopyright{acmlicensed}
\copyrightyear{2025}
\acmYear{2025}
\setcopyright{cc}
\setcctype{by}
\acmConference[SIGIR '25]{Proceedings of the 48th International ACM SIGIR Conference on Research and Development in Information Retrieval}{July 13--18,
2025}{Padua, Italy}
\acmBooktitle{Proceedings of the 48th International ACM SIGIR Conference on
Research and Development in Information Retrieval (SIGIR '25), July 13--18,
2025, Padua, Italy}\acmDOI{10.1145/3726302.3730336}
\acmISBN{979-8-4007-1592-1/2025/07}

\usepackage{url}
\usepackage{amsmath,amsfonts}
\usepackage{algorithmic}
\usepackage{graphicx}
\usepackage{textcomp}
\usepackage{xcolor}
\usepackage{hyperref}
\usepackage{caption}
\usepackage{subcaption}
\usepackage{makecell}
\usepackage{multirow}
\usepackage{multicol}
\usepackage{booktabs} 
\usepackage{balance}

\newcommand{\etal}{{\em et al.~}}
\newcommand{\eg}{{\em e.g.,~}}

\newcommand{\jiaxi}[1]{{\color{blue}[Jiaxi: #1]}}

\newcommand{\shuyi}[1]{{\color{purple}[shuyi: #1]}}

\begin{document}

\title{Unlearning for Federated Online Learning to Rank: A Reproducibility Study}

\author{Yiling Tao}
\authornote{Both authors contributed equally to this research.}
\authornote{Work done while pursuing a bachelor’s degree at SCUT.}
\affiliation{%
  \institution{Tsinghua University}
  \city{Shenzhen}
  \country{China}
}
\affiliation{%
  \institution{South China University of Technology}
  \city{Guangzhou}
  \country{China}
}
\email{202164010387@mail.scut.edu.cn}

\author{Shuyi Wang}
\authornotemark[1] 
\authornote{Corresponding Author}
\affiliation{%
  \institution{The University of Queensland}
  \city{Brisbane}
  \state{QLD}
  \country{Australia}
}
\email{shuyi.wang@uq.edu.au}

\author{Jiaxi Yang}
\authornote{Work done while pursuing a master’s degree at UESTC.}
\affiliation{%
  \institution{University of Electronic Science and Technology of China}
  \city{Chengdu}
  \country{China}
}
\email{jmy5701@psu.edu}

\author{Guido Zuccon}
\affiliation{%
 \institution{The University of Queensland}
 \city{Brisbane}
 \state{QLD}
 \country{Australia}}
\email{g.zuccon@uq.edu.au}

\renewcommand{\shortauthors}{Yiling Tao, Shuyi Wang, Jiaxi Yang, and Guido Zuccon}

\begin{abstract}
This paper reports on findings from a comparative study on the effectiveness and efficiency of federated unlearning strategies within Federated Online Learning to Rank (FOLTR), with specific attention to systematically analysing the unlearning capabilities of methods in a verifiable manner. 

Federated approaches to ranking of search results have recently garnered attention to address users privacy concerns. In FOLTR, privacy is safeguarded by collaboratively training ranking models across decentralized data sources, preserving individual user data while optimizing search results based on implicit feedback, such as clicks.

Recent legislation introduced across numerous countries is establishing the so called ``\emph{the right to be forgotten}'', according to which services based on machine learning models like those in FOLTR should provide capabilities that allow users to remove their own data from those used to train models. This has sparked the development of unlearning methods, along with evaluation practices to measure whether unlearning of a user data successfully occurred. Current evaluation practices are however often controversial, necessitating the use of multiple metrics for a more comprehensive assessment -- but previous proposals of unlearning methods only used single evaluation metrics. 

This paper addresses this limitation: our study rigorously assesses the effectiveness of unlearning strategies in managing both under-unlearning and over-unlearning scenarios using adapted, and newly proposed evaluation metrics.
Thanks to our detailed analysis, we uncover the strengths and limitations of five unlearning strategies, offering valuable insights into optimizing federated unlearning to balance data privacy and system performance within FOLTR. 
We publicly release our code and complete results at \url{https://github.com/Iris1026/Unlearning-for-FOLTR.git}.

\end{abstract}




\maketitle

\section{Introduction}

Online Learning to Rank (OLTR)~\cite{yue2009interactively,oosterhuis2018differentiable,oosterhuis2016probabilistic,hofmann2013reusing,zhuang2020counterfactual,jia2021pairrank} utilises users' queries and interactions on search results to train the ranking model in real-time, wherein a central server commonly conducts this training process. This centralization brings challenges in protecting user's privacy, as the data that captured by the server for training, i.e., user's queries and interactions in the form of clicks, can reveal private user information like demographic attributes, personal habits, or political views~\cite{barbaro2006face, DBLP:conf/iir/Carpineto016, sousa2021privacy}. To safeguard user's privacy, Federated Online Learning to Rank (FOLTR) leverages training strategies from Federated Learning (FL)~\cite{mcmahan2017communication} to perform OLTR~\cite{kharitonov2019federated, wang2021federated, wang2021effective}. 
In FOLTR, each client uses its own data to update the ranker locally, then sends the ranker update to a central server instead of sending the data itself. The server aggregates the received updates to produce an updated global ranker, which is subsequently broadcast to all clients, achieving a privacy-enhanced online training paradigm through federation. This method has been experimentally shown to deliver effective and privacy-preserving rankers~\cite{wang2021effective}.

Recent legislation approved across several countries establishes the ``\emph{the right to be forgotten}'', as exemplified by the European Union's General Data Protection Regulation (GDPR)~\cite{voigt2017eu}. This legislation prescribe that in any machine learning system, a person may request the removal of their own data from the trained model. In the context of a FOLTR system, this means that a client leaving the federation can request the removal of the contributions made to the ranker by their data. 

\begin{figure}[t!]
  \centering
    \centering
    \vspace{-10pt}
    \includegraphics[width=1\linewidth]{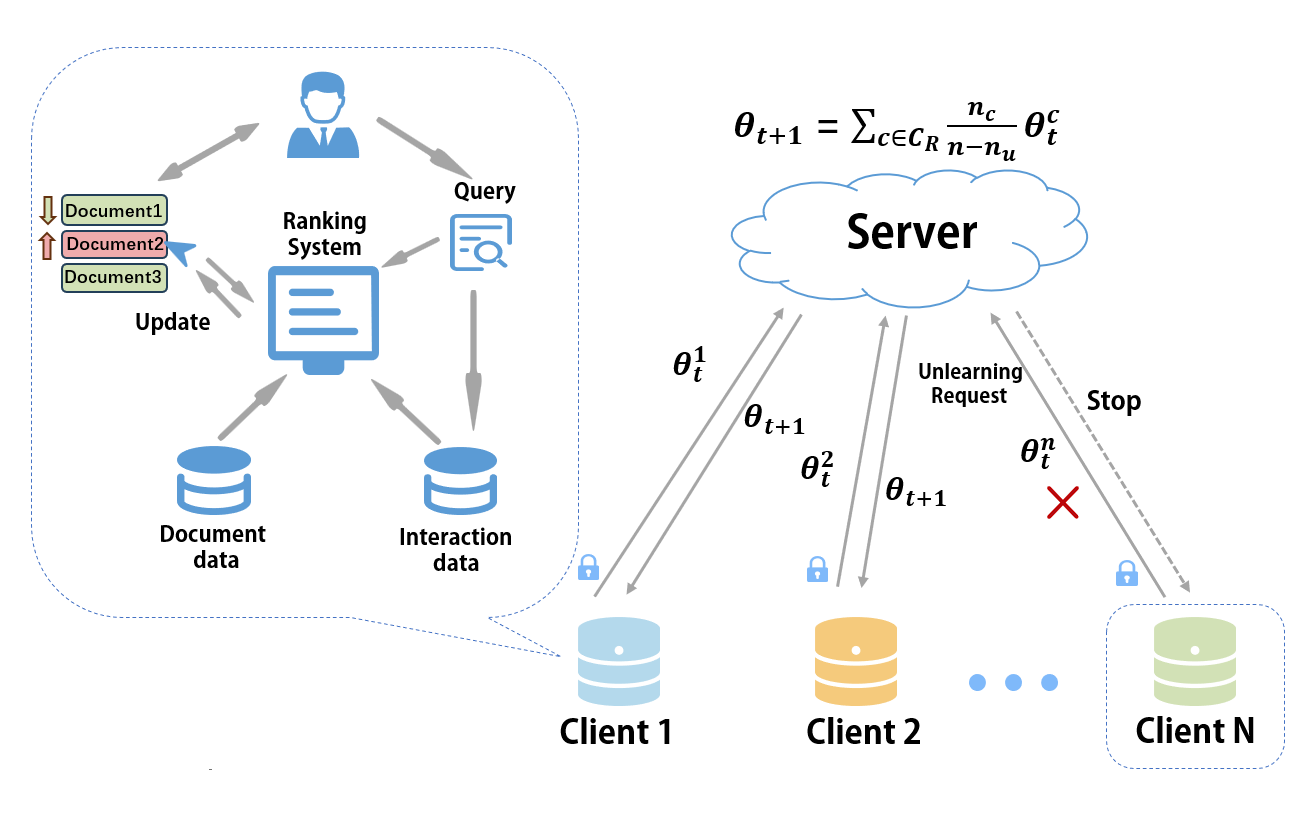}
    \vspace{-15pt}
    \caption{Unlearning in Federated Online Learning to Rank. While each client performs OLTR training locally, \(\{\theta_t^1, \theta_t^2, ..., \theta_t^n\}\) represent the local ranker parameters for client \(\{1,2,...,N\}\) separately. \( C_R \) denotes the set of remaining clients that have not request unlearning. At timestamp \(t\) of unlearning, local parameters from \( C_R \) will be sent to the server and contributes to the updated global ranker \(\theta_{t+1}\), which will be sent back to each client of \( C_R \). For clients that issue an unlearning request, their parameters will no longer be uploaded to the server, and their contributions to the server will be removed. }
    \label{fig:1} 
\end{figure}

Machine Unlearning (MU) and Federated Unlearning (FU) methods address this need, though current FOLTR systems lack effective solutions for removing specific user contributions from the global ranker, with the only method in this context being the one proposed by Wang~\etal~\cite{wang2024forget} who adapted the FedEraser strategy~\cite{liu2021federaser} to FOLTR and evaluated its unlearning effectiveness under a sophisticated simulation process. Figure~\ref{fig:1} illustrates the mechanism of Unlearning for FOLTR.
%
%
However, the initial work 
only studied a single unlearning method and used limited evaluation metrics. 
This narrow focus falls short of capturing the full complexity of unlearning within the FOLTR environment. 
Also, other types of unlearning approaches have not been studied in FOLTR, leaving their applicability unverified. 
%

In this paper, we adapt and reproduce multiple general unlearning strategies under the FOLTR experimental settings and provide a comprehensive codebase that includes implementations of these strategies and a unified evaluation pipeline, focusing on evaluating the effectiveness of unlearning, the risk of over-unlearning, and recovery efficiency. 
Unlike the previous study that focused on a single unlearning method with limited evaluation metrics, our work expands the scope of reproducibility research by integrating multiple unlearning strategies that have never been implemented in FOLTR before, along with a more comprehensive set of evaluation metrics for thorough analysis.
The main contributions of this paper are summarized as follows:
\begin{itemize}
 \item 
We reproduce five general Federated Unlearning strategies -- FedEraser~\cite{liu2021federaser}, Gradient Ascent~\cite{halimi2022federated}, a Fine Tuning-based method~\cite{zhang2023fedrecovery}, FedRemove~\cite{yuan2023federated}, and Retraining -- within the FOLTR settings, showcasing their versatility and effectiveness across four popular LTR datasets. 
\item 
We evaluate the effectiveness of each unlearning strategy in handling under-unlearning (incomplete data removal) and over-unlearning (excessive data removal), while also measuring their efficiency, particularly their recovery speed from poisoning attacks.
\item 
We release a publicly available codebase, which includes implementations of unlearning strategies and a unified evaluation pipeline to facilitate future research.
\end{itemize}

\section{Related Work}

\subsection{Federated Online Learning to Rank}
As a specialized distributed machine learning paradigm, FL enables collaborative training while preserving user privacy. To satisfy privacy-preserving requirements in OLTR, FL has also been introduced to this field termed federated online learning to rank (FOLTR). The FOLtR-ES method does this by incorporating evolution strategies to optimize the ranking model and \(\epsilon\)-local differential privacy for enhanced data protection~\cite{kharitonov2019federated}, but this methods is found to have reduced effectiveness and adaptability on large-scale datasets~\cite{wang2021federated}. A more effective approach, FPDGD~\cite{wang2021effective}, has been proposed by integrating the advanced Pairwise Differentiable Gradient Descent (PDGD)~\cite{oosterhuis2018differentiable} into the FL framework. Thanks to the superiority of locally-deployed PDGD ranker in handling noise and biases from user interactions, FPDGD has demonstrated strong ranking performance which is comparable to that of the state-of-the-art centralised OLTR methods. Following this, Wang and Zuccon~\cite{wang2022non} investigate the robustness of FPDGD to non-independent and identically distributed client data. The vulnerability of FPDGD to poisoning attacks in the federated system is also simulated and verified~\cite{wang2023analysis}, which extends the landscape of study on FOLTR.

\subsection{Federated Unlearning}

Federated Unlearning extends the principles of machine unlearning to FL settings, allowing individual clients' data to be removed from a global model. 
A naive approach in FU is to retrain the global model from scratch without involving clients who requested deletion, ensuing complete removal but at a high computational cost.
In order to accelerate the FU process, FedEraser adjusts historical updates from clients to reconstruct unlearned models~\cite{liu2021federaser}; other computation-efficient approaches have followed~\cite{wu2022federated,liu2022right,halimi2022federated}. 
While FU has been studied across natural language processing and recommendation systems~\cite{yuan2023federated,zhu2023heterogeneous}, its application to FOLTR tasks remains unexplored, with Wang et al.~\cite{wang2024forget} presenting the only approach to date. This study implemented an existing FU approach and evaluated its unlearning performance under a user simulation inspired by poisoning attacks~\cite{wang2023analysis, shejwalkar2021manipulating}. While Wang et al.~\cite{wang2024forget} pointed out that the uniqueness of FOLTR including optimising ranking tasks with implicit user feedback in an online manner, the challenge of adapting existing FU methods to FOLTR and evaluating the unlearning performance are largely unexplored. In this paper, we close this gap by evaluating within the context of FOLTR the performance of a broad range of unlearning methods across a diverse set of evaluation metrics.

\section{Unlearning for Federated Online Learning to Rank}

\subsection{Preliminary: Federated Online Learning to Rank}

In OLTR, given a user query \(q\) and a list of candidate documents, a ranking model \(f\) is leveraged to determine the final ranked list. The ranking system updates the model parameter \(\theta\) based on user interactions with the ranking list, e.g. clicks. FOLTR moves this ranker updating process into a hierarchical structure with each client deploying a local ranker trained by the OLTR algorithm and a central server aggregating all local models to produce a global ranker update -- the global ranker is then distributed back to individual clients for the next learning iteration. Formally: Let $c_i$ represent the $i^{th}$ client in FOLTR, where the set of all clients is denoted as $C = \{c_1, ..., c_n\}$. We assume that the local data\footnote{\scriptsize In OLTR, data used for training typically involves queries and corresponding candidate documents, feature representations of query-document pairs and user clicks.} of the $i^{th}$ client is denoted by \( d_i \), and the full training dataset is represented as \( D = \{d_1, ..., d_n\} \). \(\theta_t^i\) represents the local ranker parameters for $c_i$. At timestamp \(t\) of global update, the local parameters from \(C\) are sent to the server and contribute to the updated global ranker \(\theta_{t+1}\), which is then sent back to each client.

\subsection{Problem setup: Federated Unlearning}


Let \( C_{target} \) denote the set of \textit{target clients} (i.e. those leaving the federation); their data \( D_{target} \) needs to be unlearned. \( C_{remain} \) denotes the set of \textit{remaining clients} that have not undergone data erasure (i.e. have not left the federation), with their corresponding datasets represented by \( D_{remain} \). Note that $C = C_{remain} \cup C_{target}$ and $D = D_{remain} \cup D_{target}$.
The global objective function  \( \theta^{\ast} \) minimizes the empirical loss across the remaining clients after unlearning has been requested: 
\begin{equation}
\theta^\ast = \mathop{\arg\min}\limits_{\theta} \sum_{i \in C_{remain}} w_i \mathcal{L}_i(\theta, d_i).
\end{equation}
%
where the weight \( w_i \) represents the proportion of the $i^{th}$ client's dataset \(d_i\) relative to the total dataset \(D\), and \(\mathcal{L}_i(\theta,d_i) \) denotes the loss function on client \( i \)'s dataset \( d_i \) using the parameter \( \theta \).

In the \(t^{th}\) global round, the \(i^{th}\) client runs \(E\) times of OLTR update locally based on its local data \(d_i\) to optimize the model parameters \(\theta_t\) received from the previous global iteration. This results in the updated local model parameters \(\theta_{t+1}^i\). After receiving parameter updates from the \textit{remaining clients}, the server aggregates them by calculating the weighted average:
\vspace{5pt}
\begin{equation}
\label{aggregation}
\theta_{t+1} = \sum_{i \in C_{remain} } w_i \theta_{t+1}^i.
\end{equation}

The purpose of FU is to remove the contribution of the \textit{target clients} and obtain the unlearned global model \(f\) with parameters \( \theta^u \) as similar as possible to the model \(f\) with \( \theta^{\ast} \), which is trained on \( D_{remain} \) of the \textit{remaining clients}. Model similarity can be quantified by measuring the distribution similarity of model parameters. Then, we can define the problem as:
\begin{equation}
\label{distribution}
 \mathcal{T}(f_{{\theta}^{\ast}}(D_{remain})) \simeq \mathcal{T}_\mathcal{U}(f_{{\theta}^u}(D)),
\end{equation}
where \(  \mathcal{T} \) denotes the distribution of model parameters. The unlearning strategy \( \mathcal{U} \) achieves unlearning when these two distributions are as close as possible.

\subsection{Federated Unlearning Strategies}

Next we introduce the five general FU strategies we consider in this study, and their adaptation to FOLTR. 

\vspace{10pt}
\noindent{\textbf{\textit{Retraining.}}}
Retraining from scratch restarts the entire training process from the beginning with only \( C_{remain} \) and the server involved. The global model is updated based on Eq.\eqref{aggregation}. This method guarantees the complete removal of data from \textit{target clients}. While retraining is possible in standard offline classification tasks, retraining for OLTR is not possible in practice, because users cannot be asked to repeat the same interactions they have done in the past. In our study, as a baseline for comparison, we utilise retraining by repeating the same FOLTR process but not with the same historical interactions, noting its limitations\footnote{\scriptsize Retraining might still be done using offline LTR approaches like counterfactual learning; however we do not consider this possibility in this paper.}.

\vspace{10pt}
\noindent{\textbf{\textit{Fine-tuning~\cite{zhang2023fedrecovery}.}}}
Unlike retraining, FU via fine-tuning continues training of the current global model (regarded as the "pre-trained" model) once unlearning has been requested, collaboratively using interactions from the remaining clients and updating the global model based on Eq.\eqref{aggregation} (regarded as the "fine-tuning" process)~\cite{zhang2023fedrecovery}. The logic behind fine-tuning is that when the model is fine-tuned on \( D_{remain} \), it induces catastrophic forgetting on \( D_{target} \), which is a common phenomenon in continual learning processes~\cite{parisi2019continual}.

\vspace{10pt}
\noindent{\textbf{\textit{FedEraser~\cite{liu2021federaser, wang2024forget}.}}}
This method removes the influence of target clients $C_{target}$ by calibrating the updates from the remaining clients \(C_{remain}\) and aggregating these calibrated updates. Specifically, each remaining client \(i\) calibrates its local update \(\Delta \hat {\mathcal{\theta}}_{t}^i\) using the stored historical local update \(\Delta\mathcal{\theta}_{t}^i \):

\begin{equation}
\Delta\tilde{\mathcal{\theta}}_{t}^i = \frac{\Delta\hat {\mathcal{\theta}}_{t}^i}{\left\|\Delta\hat{\mathcal{\theta}}_{t}^i\right\|} {\left\|{\Delta\mathcal{\theta}}_{t}^i\right\|}.
\label{eq:calibration}
\end{equation}

\noindent
Then each client \(i\) sends the calibrated updates \(\Delta\tilde{\mathcal{\theta}}_{t}^i\) to the central server and the global model is updated: 
\begin{equation}
\theta_{t+1} = \theta_{t} + \sum_{i \in C_{remain}}w_i \Delta\tilde{\mathcal{\theta}}_{t}^i.
\end{equation}


\vspace{10pt}
\noindent{\textbf{\textit{FedRemove~\cite{yuan2023federated}.}}}
This method directly removes the parameter updates of the target clients \(C_{target}\) from the global aggregation. Same as FedEraser, before the unlearning request, the update of each client \(\Delta\mathcal{\theta}_{t}^i\) are stored after each iteration \(t\) of the federated training. When the removal happens, the stored parameters of \(C_{remain}\) are re-aggregated as the global model: 
\begin{equation}
\theta_{t+1} = \theta_{t} + \sum_{i \in C_{remain}}w_i \Delta\mathcal{\theta}_{t}^i.
\end{equation}
Compared to FedEraser, this method does not include the calibration step as Eq.\eqref{eq:calibration} and only the central server is involved and aggregates the historical local updates from \(C_{remain}\) in unlearning.
%
The advantage of this method is its rapid unlearning speed because it does not require retraining the model and it only needs to re-aggregate the updates of all clients except those being forgotten.

\vspace{10pt}
\noindent
\textbf{\textit{Gradient Ascent~\cite{halimi2022federated}.}}
This approach utilises gradient ascent to maximizes the empirical loss on \textit{target clients} for reversing the local model training process.
To prevent excessive parameter divergence, updates are projected onto an \( \ell_2 \)-norm ball centered at the reference model, which represents the average of the remaining clients' models, ensuring the unlearned model stays aligned with them. Unlike other FU strategies, this approach performs unlearning on the \textit{target clients} local side with their local interactions.
During each iteration, target client \( i \) updates its model parameters as follows:

\begin{equation}
\theta \leftarrow \mathcal{P}(\theta + \eta_u \nabla F_i(\theta))
\end{equation}

\noindent
where \( \mathcal{P} \) denotes the projection operator onto the \( \ell_2 \)-norm ball, \( \eta_u \) is the learning rate for unlearning, and \( F_i(\theta) \) represents the OLTR loss function of client \( i \).

\section{Unlearning Evaluation} 

A key challenge for unlearning lies in its evaluation practice. The goal of unlearning is to remove the contribution of target clients so as to obtain an unlearned global model. This model should be similar to a model trained from scratch on the same interactions but those from the target clients. Two types of error can lead to ineffective unlearning. When the process fails to forget the clients data, we are in presence of incomplete forgetting, or \emph{under-unlearning}. On the other hand, when the client fails to retain contributions provided by clients still in the federation, we are in presence of erroneous forgetting, or \emph{over-unlearning}. To ensure a robust evaluation of the studied unlearning strategies, in this paper we employ a broad spectrum of evaluation metrics designed to detected either  \emph{under-unlearning} or \emph{over-unlearning}~\cite{zhao2023survey}: this is the first time unlearning methods for FOLTR have been considered across both evaluation dimensions. 




\subsection{Under-unlearning}
The poisoning attack simulates malicious clients who intentionally upload manipulated updates, distorting the behaviour of the global model and degrading its overall performance. Therefore, it provides an operational scenario for evaluating the effectiveness of unlearning: if the model still remains in low effectiveness after unlearning the malicious clients, it indicates an under-unlearning issue. 

\label{under-unlearning}
\vspace{10pt}
\noindent{\textbf{\textit{Data Poisoning Attack.}}}
To measure the unlearning performance, previous studies utilize a backdoor attack which puts a marker on the target data thus the backdoor attack is only triggered by backdoor patterns in the target data~\cite{bagdasaryan2020backdoor, xie2019dba, wu2022federated}. However, this evaluation approach cannot be adapted to FOLTR directly due to the difficulties of backdoor attacks in FOLTR~\cite{wang2024forget}, as the dynamic and uncertain nature of user feedback makes it hard to maintain attack effectiveness even with injected triggers. To overcome this, we manipulate the target clients by employing a data poisoning attack~\cite{wang2023analysis} based on a reverse simulation click strategy. In particular, a client compromised to data poisoning will intentionally interact with documents unrelated to the search query. In our empirical study, we simulate the poisoned user's clicks based on the data poisoning click instantiation shown in Table~\ref{tab:table1}. This action injects misleading interaction data into the training phase and disrupts the model’s accurate prediction of user preference. The effectiveness of unlearning is thus determined by comparing the nDCG@k score before and after unlearning. Compared to the nDCG@k score before unlearning, a higher nDCG@k score achieved after unlearning is indicative of the effectiveness of the unlearning strategy.

\vspace{10pt}
\noindent{\textbf{\textit{Model Poisoning Attack.}}}
Unlike data poisoning, model poisoning directly modifies the local model updates. Previous studies have demonstrated its effectiveness in compromising global model performance and evaluating unlearning in FL systems~\cite{baruch2019little, fang2020local, shejwalkar2021manipulating, yuan2023federated, wang2024forget}. Similar to Wang et al~\cite{wang2024forget}, we also evaluate the unlearning effectiveness in the model poisoning attack scenario where we assume that the local updates \(\theta_t^i\) of \textit{target clients} have been compromised before unlearning:
\vspace{5pt}
\begin{equation}
{\theta}_t^{poi} = -\gamma {\theta}_t^i + \mu
\end{equation}

\noindent
where \({\theta}_t^{poi}\) represents the poisoned local updates  that will be sent for global aggregation. The coefficient \(\gamma\) is used to control the magnitude of these updates, while \(\mu\) directly adds noise. At each local training step before unlearning, \(\gamma\) is a random value drawn from a uniform distribution over the interval $[1, 2]$ and \(\mu\) is drawn from a normal distribution whose mean and standard deviation are determined by the original gradient updates. As these \textit{target clients} attempt to disrupt the FOLTR training process by uploading poisoned updates, the evaluation of unlearning performance has been switched to observe if the unlearned model can achieve high ranking performance after unlearning. 
In the Model Poisoning Attack, we use the same evaluation metrics as those used in the Data Poisoning Attack.\\

\vspace{10pt}
\noindent{\textbf{\textit{Relevancy Reset.}}}
In contrast to the two invasive methods mentioned above, we also introduce a non-invasive evaluation approach, Relevancy Reset (RelR), aimed at preventing performance deterioration and external security risks. Inspired by Erroneous Memory (EM)~\cite{gao2022verifi}, RelR  focuses on the samples that exhibit high loss during the model training process, which indicates the model struggles to produce a correct ranking. By paying attention to and relabeling these samples, it is possible to evaluate the unlearning effect. Specifically, we first select the top K\% high-loss queries from the datasets of a target client \(c\) under the local model \({f}_c\); we denote these as \({D}_{c}\). Then we modify the relevance labels of the documents in these queries according to the predicted ranking result by the local model \({f}_c\) and we denote the relabelled dataset as \({D}_{c}^m\) (Details of our implementation are given in Section~\ref{fed-setting}). Then we fine-tune the local model \({f}_c\) on \({D}_{c}^m\) to obtain a model \(\hat{{f}_c}\). This model is then uploaded to the central server for aggregation. After the unlearning process, an increase in the loss on the subset \({D}_{c}^m\) in the global model suggests that the model has successfully achieved the intended unlearning effect.
 

\subsection{Over-unlearning}

Another key aspect of evaluating unlearning performance is the presence of over-unlearning. Over-unlearning occurs when data that should be retained is inadvertently forgotten. To determine the occurrence of over-unlearning, we compare the alignment degree between the retrained model and the unlearned model using model performance-based and model distribution-based metrics~\cite{zhao2023survey}.

\vspace{10pt}
\noindent{\textbf{\textit{Model Performance.}}}
The straightforward approach to comparing the alignment between the unlearned model and the retrained model is through their consistency in model performance (\eg test accuracy and loss). In OLTR, we employ performance-based metric like nDCG@k for evaluation. A higher nDCG@k value indicates better unlearning performance.

 \vspace{10pt}
\noindent{\textbf{\textit{Model Distribution.}}}
An alternative approach is to evaluate the distribution of model parameters: this quantifies the alignment between the unlearned and retrained models by calculating the divergence between their distributions, using statistical measures such as \( \ell_2 \)-distance, KL divergence~\cite{goldberger2003efficient} or Wasserstein distance~\cite{panaretos2019statistical}. As shown in Eq.\eqref{distribution}, a reduced divergence indicates a higher fidelity in the unlearning process, reflecting a more effective unlearning strategy. In this paper, we quantify the parameter distribution difference between the unlearned model and the retrained model using \( \ell_2 \)-distance.

\section{Experiments}

We conduct experiments to address the following three research questions:

\begin{itemize}
 \item 
\textbf{RQ1: How do the unlearning strategies perform in terms of under-unlearning?} This question assesses the effectiveness of different FU methods in thoroughly removing the influence of target data.
 \item 
\textbf{RQ2: How do the unlearning strategies perform in terms of over-unlearning?} This explores whether the applied unlearning strategies inadvertently remove more information than necessary.
 \item 
\textbf{RQ3: What is the efficiency of the unlearning strategies?} This question evaluates the efficiency of strategies by comparing the recovery speed (from poisoning) of different unlearning strategies and the baseline \textit{Retraining} strategy.
\end{itemize}



\subsection{Experimental Settings}


\textbf{Datasets.}
We use four datasets commonly used for OLTR evaluation: MQ2007~\cite{qin2013introducing}, MSLR-WEB10K~\cite{qin2013introducing}, Yahoo~\cite{chapelle2011yahoo}, and Istella-S~\cite{lucchese2016post}. MQ2007 is the smallest with 1,700 queries. Each query-document pair is represented by a 46-dimensional feature vector, and relevance labels from \textit{not relevant} (0) to \textit{very relevant} (2). MSLR-WEB10K, Yahoo, and Istella-S are larger, and all have a 5-level relevance label scheme, from \textit{not relevant} (0) to \textit{perfectly relevant} (4). MSLR-WEB10K has 10,000 queries and 136 features, Yahoo has 29,900 queries and 700 features, and Istella-S has 33,018 queries and 220 features. MQ2007 and MSLR-WEB10K have five data splits, while Yahoo and Istella-S have a single data split. Results are averaged across all data splits.

\vspace{10pt}
\textbf{User simulations.}
As no real-world user interactions are available, we simulate user queries and click behaviours through randomly sampling N queries for each participating client in each training round and leveraging the click model for generating clicks. (This is common in FOLTR research). We employ a cascade click model (CCM)~\cite{guo2009efficient} to simulate users' click behavior. Under this click model, users examine the search engine results page from top to bottom. The CCM click model provides conditional probabilities of user clicks based on the relevance of each document. The click probability \( P(\text{click} = 1 | \text{rel}(d)) \) signifies the likelihood of a user clicking on a document related to its relevance level. The stopping probability \( P(\text{stop} = 1 | \text{click} = 1, \text{rel}(d)) \) represents the likelihood that a user decides to end their search session after clicking on a document they deem relevant. CCM has three common instances. The \textit{Perfect} user examines and clicks on all relevant documents, providing very reliable feedback. The \textit{Navigational} user may stop searching after finding the target information. They primarily focus on finding reasonably relevant documents. The \textit{Informational} user does not selectively click on documents, thus providing noisy click feedback. Table~\ref{tab:table1} illustrates the three instances of CCM and an instance simulating malicious click behaviour for Data Poisoning (Section~\ref{under-unlearning}).

\begin{table}[t]
	\centering
	\caption[centre]{Instantiations of CCM click model. \( \text{rel}(d) \) denotes the relevance label for document \( d \). The values for MQ2007 are demonstrated in brackets (only three levels of relevance are used).}
    \resizebox{0.45\textwidth}{!}{%
	\begin{tabular}{cccccc}
		\hline
            \toprule
		& \multicolumn{5}{c}{$P(\mathit{click}=1\mid rel(d))$}  \\
		\cmidrule(r){2-6}
		\emph{rel(d)} & 0 & 1 & 2 & 3 & 4 \\
		\hline
		\emph{perfect} &  0.0 (0.0)&  0.2 (0.5)&  0.4 (1.0)&  0.8 (-)&  1.0 (-)\\
		\emph{navigational} & 0.05 (0.05)& 0.3 (0.5)&  0.5 (0.95)&  0.7 (-)&  0.95 (-)\\
		\emph{informational} &  0.4 (0.4)&  0.6 (0.7)&  0.7 (0.9)&  0.8 (-)&  0.9 (-)\\
		\emph{poison} &  1.0 (1.0)&  0.8 (0.5)&  0.6 (0.0)&  0.2 (-)&  0.0 (-)\\
            \midrule
		& \multicolumn{5}{c}{$P(\mathit{stop}=1\mid click=1,  rel(d))$} \\
		\cmidrule(r){2-6}
		\emph{rel(d)} & 0 & 1 & 2 & 3 & 4 \\
		\hline
		\emph{perfect}  & 0.0 (0.0)& 0.0 (0.0)&  0.0 (0.0)&  0.0 (-)&  0.0 (-)\\
		\emph{navigational} & 0.2 (0.2)&  0.3  (0.5)&  0.5 (0.9)&  0.7 (-)&  0.9 (-)\\
		\emph{informational} & 0.1 (0.1)&  0.2 (0.3)&  0.3 (0.5)&  0.4 (-)&  0.5 (-)\\
		\emph{poison}  & 0.0 (0.0)& 0.0 (0.0)&  0.0 (0.0)&  0.0 (-)&  0.0 (-)\\
            \bottomrule
            \hline
	\end{tabular}}
	\label{tab:table1}
\end{table}

\vspace{10pt}
\textbf{Federated Settings.}
\label{fed-setting}
Our experiments involve 10 clients, each performing 5 local updates before the model update is sent to the central server. The global model is updated 1,000 times for both before unlearning and during the unlearning procedure. As ranker we employ a linear model, with a learning rate of 0.1; we train the ranker with FPDGD~\cite{wang2021effective}. When evaluating the unlearning performance under data poisoning attack and model poisoning attack, we set the poisoning rate to 30\%, with three clients subjected to a poisoning.


For the RelR metrics and datasets like MQ2007, which have three levels of relevance, we re-label the top 20\% of samples in \({D}_{c}^m\) as highly relevant (relevance=2). The subsequent 30\% are re-labeled as moderately relevant (relevance=1), and the remaining samples are re-labeled as irrelevant (relevance=0). In contrast, for datasets with five relevance levels (MSLR-WEB10K, Yahoo, and Istella-S), we re-label the top 20\% of samples in  \({D}_{c}^m\) as highly relevant (relevance=4). Each subsequent 20\% interval receives a decreasingly lower relevance score, ensuring the labels accurately reflect the predicted ranking distribution of the local model \({f}_c\).

\subsection{Results and Analysis}

\begin{figure*}[h!]
  \centering
  \begin{subfigure}[b]{0.95\textwidth}
    \centering
    \includegraphics[width=1.0\textwidth]{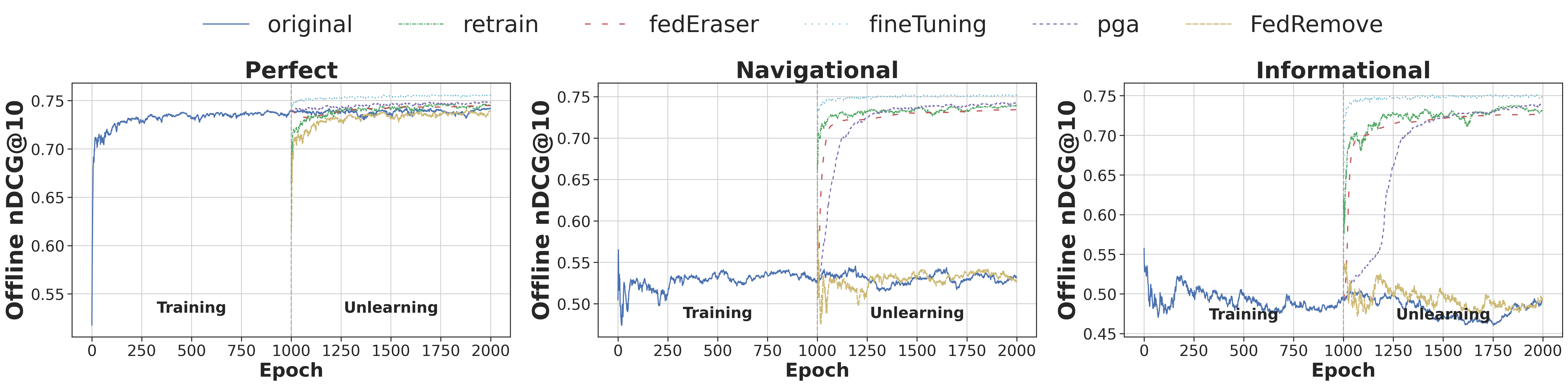}
    \caption{Data Poisoning.}
    \label{fig:2}
  \end{subfigure}
  \begin{subfigure}[b]{0.95\textwidth}
    \centering
    \includegraphics[width=1.0\textwidth]{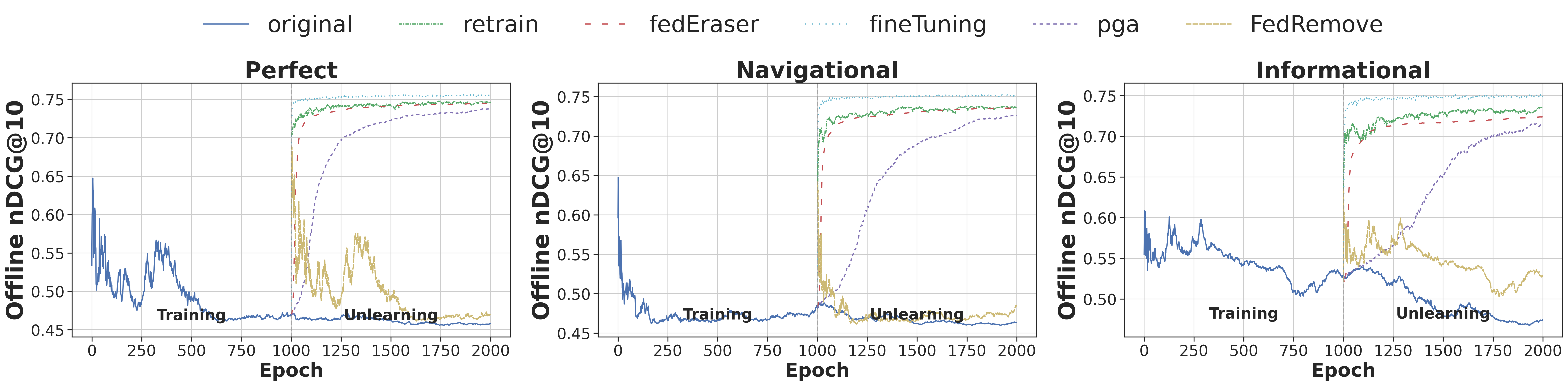}
    \caption{Model Poisoning.}
    \label{fig:3}
  \end{subfigure}
  \begin{subfigure}[b]{0.95\textwidth}
    \centering
    \includegraphics[width=1.0\textwidth]{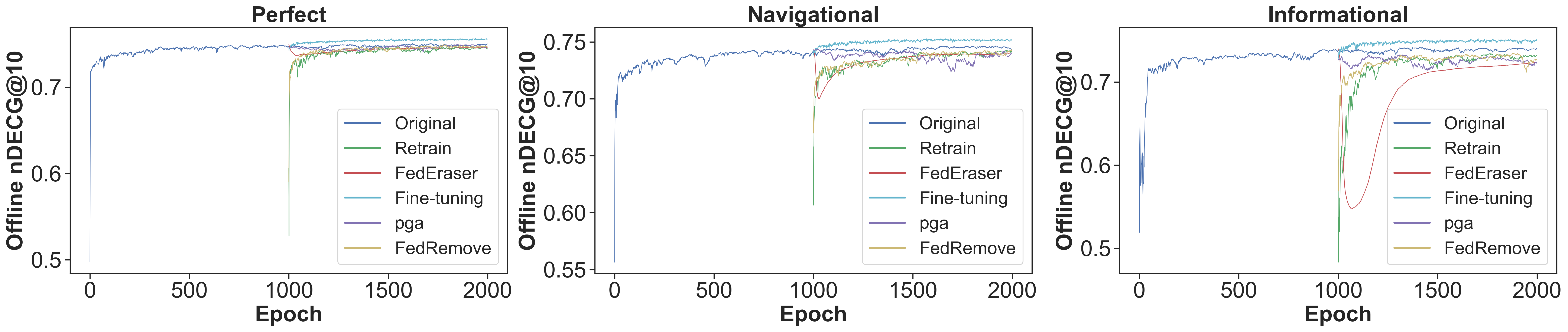}
    \caption{Clean.} 
    \label{fig:4}
  \end{subfigure}
  \caption{Offline nDCG@10 obtained under three scenarios (data poisoning, model poisoning, clean) on Yahoo dataset, under three click models (\textit{Perfect}, \textit{Navigational}, \textit{Informational}).}
  \label{xxx}
\end{figure*}

\begin{table*}[h!]
\centering
\caption{Evaluation of \textbf{Online Performance}(Discounted Cumulative NDCG@10) on four datasets, under three different click models, and averaged across dataset splits. The best results are highlighted in boldface with superscripts denoting results for statistical significance study (paired Student's t-test with $p \leq 0.05$).}
\label{table:2}
\resizebox{0.8\textwidth}{!}{%
\begin{tabular}{lcccccc}
\toprule
Dataset & Click & Poisoning   & Original(a) & Retrain(b) & FedEraser(c)  & Fine-tuning(d) \\
\midrule

\multirow{9}{*}{MQ2007} 
 & \multirow{3}{*}{Per.} 
 & Data & 548.3 & 603.3 & 616.2 & \(\textbf{618.4}^{ab}\)  \\
 & & Model & 396.7 & 488.4 & 500.9 & \(\textbf{508.0}^{ab}\)  \\
 & & None   & \(\textbf{654.3}^{c}\)  & 626.9 & 647.0 &  648.8  \\
 \cmidrule(lr){3-7}
 & \multirow{3}{*}{Nav.} 
 & Data  & 356.4 & 455.8 & 471.9 & \(\textbf{485.2}^{ab}\)  \\
 & & Model & 335.1 & 458.6 & 476.5 & \(\textbf{488.4}^{abc}\)  \\
 & & None   & \(\textbf{634.6}^{bc}\) &  597.5 &  622.7 & 630.1  \\
 \cmidrule(lr){3-7}
 & \multirow{3}{*}{Inf.} 
 & Data & 305.8 & 429.5 & 441.0 &  \(\textbf{457.3}^{ab}\)  \\
 & & Model & 381.4 & 512.1 & 524.3 & \(\textbf{540.2}^{ab}\)  \\
 & & None   & \(\textbf{631.9}^{bc}\) &  595.0 & 617.1 & 627.3  \\
 \midrule
 
 \multirow{9}{*}{MSLR-WEB10K}
 & \multirow{3}{*}{Per.} 
 & Data & 495.0 & 541.5 & 563.3 &  \(\textbf{585.5}^{abc}\) \\
 & & Model & 375.2 & 436.5 & 458.2 &  \(\textbf{479.9}^{abc}\)  \\
 & & None   & \(\textbf{621.8}^{bc}\) & 576.8 & 603.3 & 621.0  \\
 \cmidrule(lr){3-7}
 & \multirow{3}{*}{Nav.} 
 & Data & 401.8 & 452.8 & 481.9 & \(\textbf{512.7}^{abc}\)   \\
 & & Model & 398.4 & 453.3 & 482.4 & \(\textbf{512.7}^{abc}\)  \\
 & & None   & \(\textbf{601.5}^{bc}\) & 540.6 & 574.7 & 601.1  \\
 \cmidrule(lr){3-7}
 & \multirow{3}{*}{Inf.} 
 & Data & 317.2 & 371.4 & 392.0 & \(\textbf{437.0}^{abc}\)   \\
 & & Model & 382.7 & 429.6 & 449.0 & \(\textbf{495.4}^{abc}\)  \\
 & & None   & \(\textbf{589.6}^{bc}\) & 522.1 & 545.6 & 589.0  \\
  \midrule

  \multirow{9}{*}{Yahoo}
 & \multirow{3}{*}{Per.} 
 & Data & 1041.8 & 1092.9 & 1096.6 & \(\textbf{1111.9}^{a}\)   \\
 & & Model & 854.2 & 894.8 & 922.2 & \(\textbf{938.4}^{ab}\)  \\
 & & None   & \(\textbf{1124.6}^{bc}\) & 1105.3 & 1110.5 & 1124.1 \\
 \cmidrule(lr){3-7}
 & \multirow{3}{*}{Nav.} 
 & Data & 874.1 & 942.6 & 945.9 & \(\textbf{967.8}^{abc}\) \\
 & & Model & 867.5 & 898.6 & 945.5 & \(\textbf{967.8}^{abc}\) \\
 & & None   & \(\textbf{1112.8}^{bc}\) & 1086.9 & 1092.4 & 1112.4 \\
 \cmidrule(lr){3-7}
 & \multirow{3}{*}{Inf.} 
 & Data & 829.7 & 903.2 & 907.2 & \(\textbf{936.6}^{abc}\)  \\
 & & Model & 863.3 & 882.5 & 938.9 & \(\textbf{969.5}^{abc}\) \\
 & & None   & \(\textbf{1103.9}^{bc}\) & 1069.0 & 1057.7 & 1103.0 \\
  \midrule

 \multirow{9}{*}{Istella-S}
 & \multirow{3}{*}{Per.} 
 & Data & 844.9 & 972.2 & 992.9 & \(\textbf{1013.3}^{ab}\)   \\
 & & Model & 565.6 & 657.4 & 721.7 & \(\textbf{742.4}^{ab}\) \\
 & & None   & \(\textbf{1047.6}^{bc}\) & 1005.8 & 1030.2 & 1047.1 \\
 \cmidrule(lr){3-7}
 & \multirow{3}{*}{Nav.} 
 & Data & 481.9 & 606.5 & 641.4 & \(\textbf{686.8}^{abc}\) \\
 & & Model & 417.4 & 474.4 & 582.1 &  \(\textbf{627.1}^{abc}\)  \\
 & & None   & \(\textbf{992.1}^{bc}\) & 910.6 & 955.3 & 991.3 \\
 \cmidrule(lr){3-7}
 & \multirow{3}{*}{Inf.} 
 & Data & 383.0 & 503.4 & 537.5 & \(\textbf{582.2}^{abc}\)  \\
 & & Model & 374.9 & 348.4 & 533.2 &  \(\textbf{574.3}^{abc}\) \\
 & & None   & \(\textbf{993.5}^{bc}\) & 913.3 & 956.6 & 993.0 \\
  
\bottomrule

\end{tabular}
}
\end{table*}

We present a subset of the Offline nDCG@10 plots. Similar trends are observed on the other datasets but are omitted due to space constraints. Complete results are available at \url{https://github.com/Iris1026/Unlearning-for-FOLTR.git}. 

\vspace{5pt}
\textbf{RQ1: How do the unlearning strategies perform in terms of under-unlearning?}

\noindent\textbf{Experimental Design.} 
Under-unlearning evaluation aims to determine if the data contributed by the target clients has been thoroughly removed from the ranker. In this experiment, the assessment is conducted using three metrics: Relevancy Reset Difference (RelR Diff), Online performance and Offline performance. Online performance is the cumulative discounted nDCG@10 of the rankings displayed to users during training, measuring the user experience throughout the training process. For \( T \) sequential queries with \( R^t \) as the ranking displayed to the user at timestep \( t \), this is:

\begin{equation}
\text{Online Performance} = \sum_{t=1}^{T} \text{nDCG}(R^t) \cdot \gamma^{(t-1)}
\end{equation}
We follow previous work~\cite{oosterhuis2018differentiable} by choosing a discount factor of \(\gamma=0.9995\).
 Offline performance is the averaged nDCG@10 of the global ranker on the test set, showcasing the overall performance of the model. Online performance and offline performance are discussed under two types of attacks: Data Poisoning and Model Poisoning. And RelR Diff is used in the absence of attacks. 

\noindent\textbf{Results Analysis.}
Figure~\ref{fig:2} reports the offline nDCG@10 under data poisoning, while Figure~\ref{fig:3} illustrates it under model poisoning. The "Original" line represents the globally poisoned model before unlearning, with its nDCG@10 score remaining consistently low in the poisoned state. During epochs 0 to 999, each client performs federated training, followed by unlearning from epochs 1000 to 2000. We find that,  \textit{Fine-tuning} demonstrates the best unlearning effect, outperforming the \textit{Retraining} across all three click models by effectively removing the impact of target data (with a significant improvement in nDCG@10 compared to the original model). \textit{FedEraser}, and \textit{Gradient Ascent} methods perform comparably to \textit{Retraining}, occasionally even surpassing it, demonstrating strong unlearning capabilities. By contrast, \textit{FedRemove} demonstrates the poorest performance—occasionally performing even worse than the original model. This underwhelming result is primarily due to its lack of a calibration mechanism, unlike\textit{ FedEraser}, and the fact that it operates solely on the server side. As a result, \textit{FedRemove} is less effective at mitigating historical biases and counteracting adversarial updates. \\
For online performance, as shown in Table~\ref{table:2}, \textit{Fine-tuning} again leads, offering the best user experience during unlearning, followed by \textit{FedEraser} and \textit{Retraining}. Since \textit{Gradient Ascent} performs unlearning on target clients and \textit{FedRemove} operates on the server, a comparison of online performance for remaining the clients during the unlearning phase is not feasible.\\
Table~\ref{table:3} highlights RelR Diff across unlearning strategies, where \textit{FedEraser} stands out with the highest value, underscoring its unlearning effectiveness.\\

\textbf{RQ2: How do the unlearning strategies perform in terms of over-unlearning?}

\noindent \textbf{Experimental Design.} Another contrasting evaluation perspective to under-unlearning in FU is over-unlearning, which assesses whether essential data that does not require unlearning, apart from the target clients' data, has also been inadvertently removed.
In this experiment, we investigate the prevalence of over-unlearning by analyzing how closely various unlearning strategies align with \textit{Retraining} outcomes. We utilize three metrics: Distance Difference (Dist Diff), Online performance(without poisoning) and Offline performance(without poisoning). Dist Diff quantifies the parameter distribution difference between the unlearned model and the retrained model (calculated using \( \ell_2 \)-distance). A close alignment with \textit{Retraining} on Dist Diff suggests better unlearning performance in terms of over-unlearning, while a higher nDCG@10 value generally indicates better ranking performance of the unlearned model.

\begin{figure*}[t!]
  \centering
    \centering
    \includegraphics[width=1\linewidth]{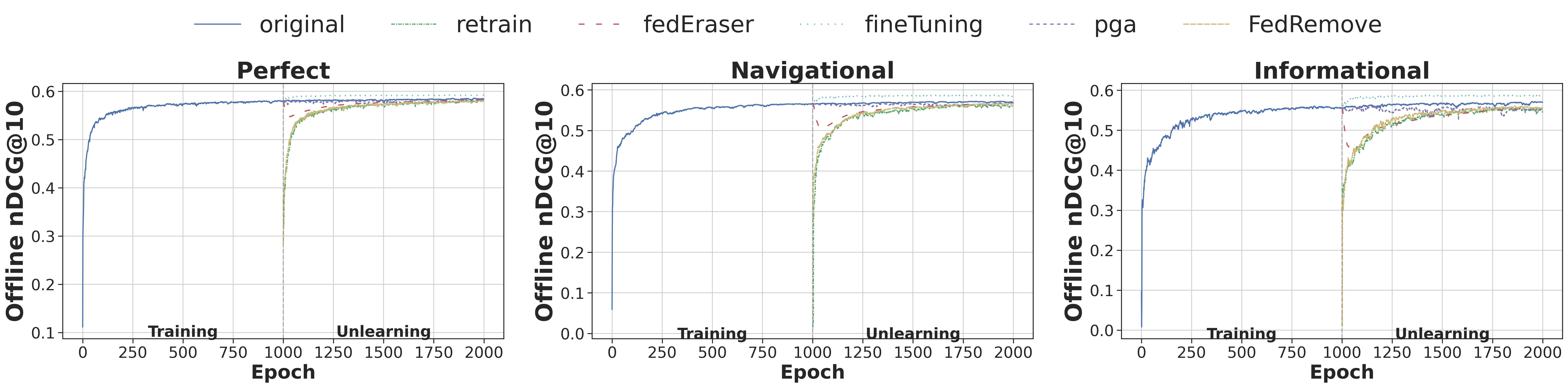}
    \caption{ Offline nDCG@10 obtained under clean scenario on Istella-S dataset, under three click models (Perfect,
Navigational, Informational).}
    \label{fig:5} 
\end{figure*}

\noindent\textbf{Results Analysis.} 
In the evaluation of over-unlearning, different unlearning strategies exhibit subtle differences, particularly for the Dist Diff metric (shown in Table~\ref{table:3}). \textit{Gradient Ascent} shows a significantly higher distance compared to other strategies, indicating its poorer ability to maintain model integrity after unlearning, leading to over-unlearning. \textit{FedEraser}, \textit{Gradient Ascent}, and \textit{Fine-tuning} all demonstrate stable model accuracy (with Offline nDCG@10, as shown in Figure~\ref{fig:4} and Figure~\ref{fig:5} , comparable to Retraining). However, \textit{FedRemove} performs the worst in terms of both model integrity and accuracy.\\
In terms of online performance (shown in Table~\ref{table:2}), without a poisoning attack, \textit{Original} achieves the best online performance. Other strategies, such as \textit{FedEraser}, \textit{Fine-tuning}, and \textit{Retrain}, perform slightly worse than \textit{Original} but still manage to maintain user experience during the unlearning process. For the same reasons as in RQ1, we do not compare \textit{Gradient Ascent} and \textit{FedRemove}.\\

\begin{table*}[h!]
\centering
\caption{RelR Difference(RelR Diff) and \( \ell_2 \)-Distance Difference(Dist Diff) on four datasets, under three different click models, and averaged across dataset splits. The best results are highlighted in boldface with superscripts denoting results for statistical significance study (paired Student's t-test with $p \leq 0.05$).}
\label{table:3}
\resizebox{0.9\textwidth}{!}{%
\begin{tabular}{lccccccc}
\toprule
Dataset & Model & Metrics  & Retrain(a) & FedEraser(b) & Gradient Ascent(c) & Fine-tuning(d) & FedRemove(e) \\
\midrule

\multirow{6}{*}{MQ2007} 
& \multirow{2}{*}{Per.} 
 & RelR Diff  & 1.50 & \(\textbf{17.07}^{acde}\) & 0.83 & 3.07 & 0.01\\
 & & Dist Diff  & - & 6.67 & 118.76 &\(\textbf{1.09}^{bce}\)&  6.24 \\
 \cmidrule(lr){3-8}
 & \multirow{2}{*}{Nav.} 
 & RelR Diff  & 1.29 & \(\textbf{17.45}^{acde}\) & 0.43 & 1.97 & 0.01 \\
 & & Dist Diff & -& 8.30 & 69.23 & \(\textbf{1.86}^{bce}\) & 7.26 \\
 \cmidrule(lr){3-8}
 & \multirow{2}{*}{Inf.} 
 & RelR Diff  & 1.35 & \(\textbf{17.25}^{acde}\) & 0.40 & 2.21 & -0.01 \\
 & & Dist Diff &- & 8.84 & 78.80 & \(\textbf{3.15}^{bce}\) & 7.47 \\
 \midrule
 
 \multirow{6}{*}{MSLR-WEB10K}
 & \multirow{2}{*}{Per.} 
 & RelR Diff  & 2.21 & \(\textbf{16.31}^{ace}\) & 0.85 & 11.22 & 0.04\\
 & & Dist Diff & - & 4.28 & 144.11 &  \(\textbf{2.04}^{bce}\)  &  5.65\\
 \cmidrule(lr){3-8}
 & \multirow{2}{*}{Nav.} 
 & RelR Diff & 2.97 & \(\textbf{16.74}^{acde}\) & 0.62 & 10.59 & -0.02 \\
 & & Dist Diff & - & 4.46 & 92.32 &  \(\textbf{2.02}^{bce}\) &  7.63\\
 \cmidrule(lr){3-8}
 & \multirow{2}{*}{Inf.} 
 & RelR Diff & 2.55  & \(\textbf{17.43}^{acde}\) & 0.90 & 11.66 & 0.07 \\
 & & Dist Diff & - & 6.55 & 108.76 & \(\textbf{3.82}^{bce}\) &  8.05\\
  \midrule

  \multirow{6}{*}{Yahoo}
 & \multirow{2}{*}{Per.} 
 & RelR Diff & 6.09 & \(\textbf{14.34}^{ace}\)  & 0.95 & 10.17 & -0.17 \\
 & & Dist Diff  & - & 5.87 &  149.75 & \(\textbf{4.00}^{bce}\) &  5.91 \\
 \cmidrule(lr){3-8}
 & \multirow{2}{*}{Nav.} 
 & RelR Diff & 4.31 & \(\textbf{15.16}^{acde}\) & 0.87 & 9.55 & 0.04\\
 & & Dist Diff & - &  7.86 & 99.43 & \(\textbf{4.80}^{bce}\) &  9.05\\
 \cmidrule(lr){3-8}
 & \multirow{2}{*}{Inf.} 
 & RelR Diff & 5.26 & \(\textbf{15.04}^{ace}\) & 0.82 & 10.64 & -0.02\\
 & & Dist Diff & - & 10.25 & 102.08 & \(\textbf{6.13}^{bce}\) &  10.41\\
  \midrule

 \multirow{6}{*}{Istella-S}
 & \multirow{2}{*}{Per.} 
 & RelR Diff  & 4.21 & \(\textbf{15.48}^{ace}\) & 0.84 & 10.72 & 0.05\\
 & & Dist Diff  & - & 5.16 & 117.80 & \(\textbf{1.28}^{bce}\) &  6.92 \\
 \cmidrule(lr){3-8}
 & \multirow{2}{*}{Nav.} 
 & RelR Diff & 2.90 & \(\textbf{16.57}^{acde}\) & 0.12 & 7.49 & 0.03\\
 & & Dist Diff  & - & 9.35 & 94.19 & \(\textbf{1.60}^{bce}\) &  6.59\\
 \cmidrule(lr){3-8}
 & \multirow{2}{*}{Inf.} 
 & RelR Diff  & 3.27 &  \(\textbf{16.09}^{acde}\) & 0.19 & 8.72 & 0.06\\
 & & Dist Diff  &- & 8.73 & 117.42 & \(\textbf{3.24}^{bce}\) &  9.59\\
  
\bottomrule
\end{tabular}
}
\end{table*}

\textbf{RQ3: What is the efficiency of the unlearning strategies?}

\noindent \textbf{Experimental Design.} To evaluate the efficiency of different unlearning strategies, we compare the recovery speed (from data poisoning and model poisoning) for each method. In this experiment, Retraining serves as the baseline. If an unlearning strategy achieves better recovery than Retraining within the same number of global training steps, it indicates that this strategy has a faster recovery speed.\\
\noindent \textbf{Results Analysis.}
From Figure~\ref{fig:2} and~\ref{fig:3} , we find that \textit{Fine-tuning} reaches convergence first, indicating that the \textit{Fine-tuning} strategy has higher recovery efficiency, as it approaches its maximum recovery level with fewer training steps. \textit{FedEraser} and \textit{Gradient Ascent} have a convergence speed comparable to \textit{Retraining}. The \textit{FedRemove} curve shows significant fluctuations across all datasets and never stabilizes, indicating that it has not effectively converged.

\section{Conclusion}

Research on unlearning methods for FOLTR remains in its early stages. To date, limited knowledge is available due to lacking empirical studies across unlearning approaches as well as a not well-established evaluation practice that does not consider a variety of important metrics. 
To gain more insight into unlearning approaches for FOLTR,
we reproduce five general federated unlearning strategies and assess them across four popular LTR datasets, applying a range of evaluation metrics to capture the nuanced performance of each strategy. 

We find that, for under-unlearning (RQ1), \textit{Fine-tuning} performs best under data and model poisoning conditions, with \textit{FedEraser} and \textit{Gradient Ascent} generally maintaining performance comparable to \textit{Retraining}, while \textit{FedRemove} shows the weakest unlearning effectiveness. In terms of over-unlearning (RQ2), \textit{Fine-tuning} effectively retains non-target data, \textit{FedEraser} performs well, whereas \textit{Gradient Ascent} tends to over-unlearn. Regarding efficiency (RQ3), \textit{Fine-tuning} demonstrates a significantly faster recovery speed (from poisoning) than other strategies, with \textit{Gradient Ascent} and \textit{FedEraser} maintaining efficiency comparable to \textit{Retraining}. 

Overall, this paper provides a comparative analysis of five FU strategies within FOLTR, revealing key insights into differences in data retention accuracy and recovery efficiency across strategies. These findings provide foundational support for optimizing FU strategies to balance privacy and performance in FOLTR.

%

\vfill\eject
\balance
\bibliographystyle{ACM-Reference-Format}
\bibliography{reference}

\end{document}